\documentclass[journal,onecolumn,11pt]{IEEEtran}

\usepackage{ifpdf}

\usepackage{cite}
\ifCLASSOPTIONcompsoc
\usepackage[nocompress]{cite}
\else
\usepackage{cite}
\fi

\ifCLASSINFOpdf
   \usepackage[pdftex]{graphicx}
   \DeclareGraphicsExtensions{.pdf,.jpeg,.png}
\else
   \usepackage[dvips]{graphicx}
   \DeclareGraphicsExtensions{.eps}
\fi

\usepackage[cmex10]{amsmath}

\usepackage[table]{xcolor}
\usepackage{amssymb}
\usepackage{stfloats}
\usepackage{amssymb}
\usepackage{bm} 
\usepackage{amsmath} 
\usepackage{subfigure}
\usepackage{bm} 

\usepackage{array}

\ifCLASSOPTIONcaptionsoff
  \usepackage[nomarkers]{endfloat}
 \let\MYoriglatexcaption\caption
 \renewcommand{\caption}[2][\relax]{\MYoriglatexcaption[#2]{#2}}
\fi

\def\ba{{\mathbf{a}}}
\def\bb{{\mathbf{b}}}
\def\bg{{\mathbf{g}}}
\def\bh{{\mathbf{h}}}
\def\bx{{\mathbf{x}}}

\def\bw{{\mathbf{w}}}

\def\bPhi{{\boldsymbol{S}}}
\def\bPsi{{\boldsymbol{P}}}
\def\bphi{\pmb{\phi}}

\newcommand{\E}[1]{\mathbb{E}\left[ #1 \right]}

\usepackage{url}

\begin{document}

%
\title{A Distributed Algorithm for Training Augmented Complex Adaptive IIR Filters}

\author{Azam Khalili (Corresponding Author),
        Reza~G.~Rahmati,
        ~Amir Rastegarnia
        and~Wael M. Bazzi
\thanks{A. Khalili, R. G. Rahmati, and A. Rastegarnia are with the Department of Electrical Engineering, Malayer University, Malayer,
Iran, e-mail: (a.khalili@ieee.org).}

\thanks{W. M. Bazzi is with Electrical Engineering Department, American University in Dubai, e-mail: wbazzi@aud.edu}

\thanks{Manuscript received 2016; revised  2016.}}

%

\markboth{Draft}%
{Shell \MakeLowercase{\textit{et al.}}: Bare Demo of IEEEtran.cls for Journals}

\maketitle

\begin{abstract}
In this paper we consider the problem of decentralized (distributed) adaptive learning, where the aim of the network is to train the coefficients of a widely linear autoregressive moving average (ARMA) model by measurements collected by the nodes. Such a
problem arises in many sensor network-based applications such as target tracking, fast rerouting, data reduction and data aggregation. We assume that each
node of the network uses the augmented complex adaptive infinite impulse response (ACAIIR) filter as the learning
rule, and nodes interact with each other under an incremental mode of cooperation. Since the proposed algorithm (incremental augmented complex IIR (IACA-IIR) algorithm) relies on the augmented complex statistics, it can be used to model both types of complex-valued signals (proper and improper signals). To evaluate the performance of the proposed algorithm, we use both synthetic and real-world complex signals in our simulations. The results
exhibit superior performance of the proposed algorithm over the non-cooperative ACAIIR algorithm.

\end{abstract}

\begin{IEEEkeywords}
Adaptive filter, widely linear modeling, complex signals, IIR.
\end{IEEEkeywords}

 \ifCLASSOPTIONpeerreview
 \begin{center} \bfseries EDICS Category: 3-BBND \end{center}
 \fi
%
\IEEEpeerreviewmaketitle

\section{Introduction}
Complex-valued adaptive filters appear in different applications such as communications, power systems, biomedical signal processing and sensor networks \cite{1,34,42,43,44,45,47}. One way to extract a complex-valued adaptive filter algorithm is to extend its real-valued counterpart. The obtained algorithm in this way is suitable only for proper complex-valued signals, since it relies only on the second order statistics, given by the covariance matrix $C_{{\bf xx}}=\E{{\bf x}{{\bf x}}^H}$. By definition, the term circular refers to a complex signal where its probability distribution is rotation-invariant in the complex plane. In addition,  the propriety (or second order circularity) implies the second order statistical properties of complex signals \cite{2}. In most real world applications, complex signals are second order noncircular or improper. In order to extract all the available second order information, beside the covariance matrix 
$C_{{\bf xx}}=\E{{\bf x}{{\bf x}}^H}$, we have to use the pseudocovariance matrix $P_{{\bf xx}}=\E{{\bf x}{{\bf x}}^T}$. To this end, the augmented representation, i.e., ${\bf x}^a=[{\bf x}^T{\bf x}^H]^T$  can be used to model the second-order statistical information within the complex domain \cite{3,4}. The augmented covariance matrix $C_{{\bf xx}}^a$ is then given by
\setlength{\arraycolsep}{2pt}
\begin{equation}
\label{eqn_1}
C_{{\bf xx}}^a=\E {{\bf x}^a{{\bf x}^a}^H}=
\left[ {\begin{array}{*{2}{c}}
C_{{\bf xx}} & P_{{\bf xx}} \\
P_{{\bf xx}}^*&C_{{\bf xx}}^*
 \end{array}} \right]
\end{equation}

Using the augmented representation, it is possible to design widely-linear adaptive algorithms that are able to process proper and improper complex signals \cite{35}. Different complex-valued adaptive filters that rely on the augmented complex statistics have been introduced in the literature such as the widely linear LMS (WL-LMS) \cite{37}, augmented CLMS (ACLMS) \cite{38}, augmented affine projection algorithm (AAPA) \cite{39}, widely linear recursive least squares (WL-RLS) \cite{40}, regularized normalized augmented complex LMS (RN-ACLMS) \cite{41}, and augmented complex adaptive IIR (ACAIIR) \cite{14}.

In this paper we deal with the problem of distributed adaptive learning, where a set of nodes are deployed to collaboratively estimate the coefficients
of a widely linear autoregressive moving average (ARMA) model. Such a problem arises in many sensor network-based applications such as  target tracking \cite{21,22}, fast rerouting \cite{23,49}, data reduction \cite{24} and data aggregation \cite{25,48}. To develop our proposed algorithm, we assume that each node 
uses the ACAIIR filter as the learning rule, and nodes interact with each other under an incremental mode of cooperation. In such a cooperation mode, there exists a cyclic path (Hamilton cycle) among the nods, in which the information goes through the nodes in one direction, i.e. each node passes the information to its adjacent node in a pre-determined direction \cite{15,17,28,29,30,50,51,52}\footnote{Other cooperation modes, such as diffusion is also possible, where each node communicates with all of its neighbors \cite{27,31,32,33}. In our future work, we will consider such a cooperation mode.}. To derive the proposed algorithm  (incremental augmented complex IIR (IACA-IIR)), we firstly formulate the distributed adaptive learning problem as an unconstrained minimization problem. Then, we use  stochastic gradient optimization argument and $\mathbb{CR}$ calculus framework \cite{5} (also known as Wirtinger calculus) to derive a distributed, recursive algorithm to train the complex-valued adaptive IIR filter. We use synthetic complex signal and real-world complex signal in our simulation, where the results reveal that the superior performance of the proposed algorithm, in comparison with the non-cooperative ssolution.

The rest of the paper is organized as follows: in Section II, the IACA-IIR algorithm is derived. In Section III, simulations are presented, and Section IV concludes the paper.

\textit{Notation}: We adopt small boldface letters for vectors and bold capital letters for matrices. The following
notations are adopted: $(\cdot)^*$ denotes the complex conjugate, $(\cdot)^T$ the transpose of a vector or a matrix, $(\cdot)^H$ the Hermitian transpose of a vector or a matrix and $\E{\cdot}$ represents the statistical expectation. 

\section{Derivation of IACA-IIR Algorithm}
Consider an incremental network with $L$ nodes, where each node $k$ has access to data $\{u_k (n),{\bf x}_k (n)\}$ at time $n$, where ${\bf x}_k (n)$ denotes the input vector and  
\begin{align} \label{eqn_2}
u_k(n)&=d_k(n)+v_k (n) 
\end{align}
where $v_k (n)$ denotes the noise term which is assumed as doubly white noise process with variance $\sigma_{v,k}^2$. We assume that $v_k (n_1)$ is independent of the input vector $\bx_l(n_2)$ for all $k \neq l$ and $n_1 \neq n_2$. Moreover, $d_k(n)$ (desired signal) is given by the following  widely linear model ARMA
\begin{align}
\label{eqn_2n}
d_k(n)&=\sum_{\ell=1}^M a^o_{\ell}(n)d(n-\ell) 
+\sum_{\ell=0}^N b^o_{\ell}(n)x_k(n-\ell)      
+\sum_{\ell=1}^M g^o_{\ell}(n)d_k^*(n-\ell)   
+ \sum_{\ell=0}^N h^o_{\ell}(n)x_k^*(n-\ell) 
\end{align}
where $M>0, \ N>0$ and
\begin{align}
\label{eqn_2nn}
\ba^o=\left[ {\begin{array}{*{2}{c}}
a_1  \\
a_2  \\
\vdots  \\
a_M
 \end{array}} \right], \ \ 
\bb^o=\left[ {\begin{array}{*{2}{c}}
b_0  \\
b_1  \\
\vdots  \\
b_N
 \end{array}} \right], \ \
 \bg^o=\left[ {\begin{array}{*{2}{c}}
g_1  \\
g_2  \\
\vdots  \\
g_M
 \end{array}} \right], \ \
 \bh^o=\left[ {\begin{array}{*{2}{c}}
h_0  \\
h_1  \\
\vdots  \\
h_N
 \end{array}} \right],
\end{align}
are unknown fixed weights. The coefficients of model in \eqref{eqn_2n} can be given at every node $k$ as the output of the following widely linear IIR filter    \cite{14}
\setlength{\arraycolsep}{0.0em}
\begin{align}
\label{eqn_4}
y_k(n)&=\sum_{m=1}^M a_{m,k}(n)y_k(n-m) 
 +\sum_{m=0}^N b_{m,k}(n)x_k(n-m)      
  +\sum_{m=1}^M g_{m,k}(n)y_k^*(n-m)    
  + \sum_{m=0}^N h_{m,k}(n)x_k^*(n-m)      
\end{align}
where $\{a_m, b_m, g_m, h_m\}$ are complex-valued filter coefficients. To estimate the desired response $d_k(n)$ (or equivalently, to train of the complex-valued ARMA model \eqref{eqn_4}), we can pose the following minimization problem
\begin{equation} \label{eqn_8}
 \mathop{\arg \min } \limits_{\{a_m, b_m, g_m, h_m\}} J(n)  \ \ \mathrm{with}\ \ J(n) \triangleq \sum_{k=1}^L |e_k (n)|^2=\sum_{k=1}^L e_k(n)e_k(n)^* \end{equation}
where the corresponding output error at node $k$ is given by
\begin{equation}
\label{eqn_5}
e_k (n)=d_k (n)-y_k (n)
\end{equation} 
Obviously the cost function \eqref{eqn_8} can be decomposed as $J(n)=\sum_{k=1}^L J_k(n)$ where $J_k(n)=\left|d_k (n)-y_k (n)\right|^2$.
Using the traditional iterative steepest-descent algorithm to solve \eqref{eqn_8} gives 
\begin{equation}    \label{sds}
\bphi(n)=\bphi(n-1)-\mu \sum_{k=1}^L [\nabla_{\bphi(n-1)} J_k (n)]^*
\end{equation} 
where $\mu>0$ is the step-size parameter, $\bphi(n)$ is a global estimate for $\bw^o=[\ba^o,\bg^o,\bb^o,\bh^o]^T$ at iteration $n$, and $\nabla_{\bphi(n-1)}$ is the gradient vector of $J_k (n)$ with respect to $\bphi(n-1)$.
Obviously \eqref{sds} is not a distributed 
solution as it requires global information $\bphi(n)$. To resolve this difficulty, we first introduce the following equivalent implementation for \eqref{sds} as
\begin{equation}
\label{eqval}
\left\{ \begin{array}{l}
\bw_1(0)=\bphi(n-1)       \\
\bw_k (n)=\bw_{k-1} (n)-\mu_k [\nabla_{\bphi(n-1)} J_k (n)]^*,k=1,2,...,L    \\
\bphi(n)=\bw_L (n)
\end{array} \right.
\end{equation}
where $\bw_k (n)$ the local estimate of $\bw^o$ at node $k$ and time $n$ \footnote{Note that 
$\bw_k (n)$ is compact form of the filter weights which is given by
\setlength{\arraycolsep}{0.0em}
\begin{align}
\label{eqn_9}
\bw_k (n)&=[a_{1,k}(n),\cdots, a_{M,k} (n), g_{1,k}(n),\cdots, g_{M,k}(n),b_{0,k}(n),\cdots, b_{N,k}(n),h_{1,k} (n),\cdots, h_{N,k}(n)]^T
\end{align}. }.
Due to incremental cooperation mode, node $k$ has access to $\bw_{k-1}(n)$, i.e. an estimate of $\bw^o$ at previous node $k-1$. Thus, to derive a fully distributed solution for \eqref{sds}, we use concept of incremental gradient algorithm \cite{46} and rewrite \eqref{eqval} as
\begin{equation}
\label{eqn_3}
\left\{ \begin{array}{l}
\bw_1(0)=\bphi(n-1)       \\
\bw_k (n)=\bw_{k-1} (n)-\mu_k [\nabla_{\bw_{k-1}(n)} J_k (n)]^*,k=1,2,...,L    \\
\bphi(n)=\bw_L (n)
\end{array} \right.
\end{equation}
The gradient term is given by
\setlength{\arraycolsep}{0.0em}
\begin{eqnarray}
\label{eqn_10}
\left[\nabla_{\bw_{k-1}} J_k (n)\right]^*&{}={}&-\left[e_k (n)\frac{\partial y_k^* (n)}{\partial \bw_{k-1}^* (n)}+\frac{\partial y_k (n)}{\partial \bw_{k-1}^* (n)} e_k^* (n) \right]\nonumber\\&{}={}&-\left[e_k (n) \bPhi_{\bw_k} (n)+\bPsi_{\bw_k} (n) e_k^* (n)\right]
\end{eqnarray}
where the sensitivities $\bPhi_{\bw_k} (n)$ and $\bPsi_{\bw_k} (n)$ are defined as
\begin{equation}
\label{eqn_11}
\bPhi_{\bw_k}(n)=\frac{\partial y_k^* (n)}{\partial \bw_{k-1}^* (n)}\Big|_{\bw_{k-1} (n)={\mbox{\footnotesize constant}}}
\end{equation}
\begin{equation}
\label{eqn_12}
\bPsi_{\bw_k}(n)=\frac{\partial y_k (n)}{\partial \bw_{k-1}^* (n)}\Big|_{\bw_{k-1} (n)={\mbox{\footnotesize constant}}}
\end{equation}
where
\begin{align}
\label{eqn_13}
\bPhi_{\bw_k}(n)&=[\Phi_{a_{1,k}}(n),...,\Phi_{a_{M,k}}(n),\Phi_{g_{1,k}}(n),...,\Phi_{g_{M,k}}(n)
,\Phi_{b_{0,k}}(n),...\Phi_{b_{N,k}}(n),\Phi_{h_{0,k}}(n),...,\Phi_{h_{N,k}}(n)]^T
\\
\bPsi_{\bw_k}(n)&=[\Psi_{a_{1,k}}(n),...,\Psi_{a_{M,k}}(n),\Psi_{g_{1,k}}(n),...,\Psi_{g_{M,k}}(n) ,\Psi_{b_{0,k}}(n),...\Psi_{b_{N,k}}(n),\Psi_{h_{0,k}}(n),...,\Psi_{h_{N,k}}(n)]^T          \label{eqn_14}
\end{align}
We can calculate the gradient in \eqref{eqn_10} separately. For instance, for $a_{m,k-1}$ we have
\begin{align}
\frac{\partial y_k^* (n)}{\partial a_{m,k-1}^* (n)}&=y_k^* (n-m)+\sum\limits_{l=1}^M a_{l,k-1}^*(n)\frac{\partial y_k^* (n-l)}{\partial a_{m,k-1}^* (n)}\nonumber   \\
& \hspace{1cm} +\sum\limits_{l=1}^M g_{l,k-1}^*(n)\frac{\partial y_k (n-l)}{\partial a_{m,k-1}^* (n)}  \label{eqn_15}
\\  
\frac{\partial y_k (n)}{\partial a_{m,k-1}^* (n)}&=\sum\limits_{l=1}^M a_{l,k-1}(n)\frac{\partial y_k (n-l)}{\partial a_{m,k-1}^* (n)}\nonumber  \\
& \hspace{1cm}  +\sum\limits_{l=1}^M g_{l,k-1}(n)\frac{\partial y_k^* (n-l)}{\partial a_{m,k-1}^* (n)}  \label{eqn_16}
\end{align}
As we can see from (\ref{eqn_15}) and (\ref{eqn_16}), to calculate the gradient in \eqref{eqn_10} we need to compute the derivatives $\frac{\partial y_k^* (n-l)}{\partial a_{m,k-1}^* (n)}$ for $l=1,2,\cdots,M$, which is not possible. To circumvent this issue, we consider the following approximation for a small convergence rate $\mu$ \cite{18}:
\begin{align}
\label{eqn_17}
\bw_k (n)& \approx \bw_k (n-1)\approx...\approx \bw_k (n-\tau)
\end{align}
where $\tau=max\{M,N+1\}$. Thus, the relations (\ref{eqn_15}) and (\ref{eqn_16}) can be approximated as follows
\begin{align}
\frac{\partial y_k^* (n)}{\partial a_{m,k-1}^* (n)}&=y_k^* (n-m)+\sum\limits_{l=1}^M a_{l,k-1}^*(n)\frac{\partial y_k^* (n-l)}{\partial a_{m,k-1}^* (n-l)}\nonumber \\
& \hspace{1cm}  +\sum\limits_{l=1}^M g_{l,k-1}^*(n)\frac{\partial y_k (n-l)}{\partial a_{m,k-1}^* (n-l)}  \label{eqn_18}
\\
\frac{\partial y_k (n)}{\partial a_{m,k-1}^* (n)}&=\sum\limits_{l=1}^M a_{l,k-1}(n)\frac{\partial y_k (n-l)}{\partial a_{m,k-1}^* (n-l)}\nonumber  \\
& \hspace{1cm}  +\sum\limits_{l=1}^M g_{l,k-1}(n)\frac{\partial y_k^* (n-l)}{\partial a_{m,k-1}^* (n-l)}  \label{eqn_19}
\end{align}
Since delayed version of sensitivities appearing in the  right hand  side of above relations, the sensitivities can be written as
\begin{align}
\Phi_{a_{m,k}}(n)&=y_k^* (n-m)+\sum\limits_{l=1}^M a_{l,k-1}^*(n)\Phi_{a_{m,k}}(n-l)   \nonumber   \\
& \hspace{1cm}  +\sum\limits_{l=1}^M g_{l,k-1}^*(n)\Psi_{a_{m,k}}(n-l)   \label{eqn_20}   
\\
\Psi_{a_{m,k}}(n)&=\sum\limits_{l=1}^M a_{l,k-1}(n)\Psi_{a_{m,k}}(n-l)    \nonumber  \\
& \hspace{1cm}  +\sum\limits_{l=1}^M g_{l,k-1}(n)\Phi_{a_{m,k}}(n-l) \label{eqn_21}
\end{align}
Similarly, update relations for other  sensitivities can be obtained as
\begin{align}
\Phi_{g_{m,k}}(n) &=y_k (n-m)+\sum\limits_{l=1}^M a_{l,k-1}^*(n)\Phi_{g_{m,k}}(n-l)   \nonumber  \\
& \hspace{1.2cm}  +\sum\limits_{l=1}^M g_{l,k-1}^*(n)\Psi_{g_{m,k}}(n-l)     \label{eqn_22} 
\\
\Phi_{b_{m,k}}(n)& =x_k^* (n-m)+\sum\limits_{l=1}^M a_{l,k-1}^*(n)\Phi_{b_{m,k}}(n-l)  \nonumber  \\
& \hspace{1.2cm}  +\sum\limits_{l=1}^M g_{l,k-1}^*(n)\Psi_{b_{m,k}}(n-l)  \label{eqn_23}
\\ 
\Phi_{h_{m,k}}(n)&=x_k (n-m)+\sum\limits_{l=1}^M a_{l,k-1}^*(n)\Phi_{h_{m,k}}(n-l)     \nonumber  \\
& \hspace{1.2cm}  +\sum\limits_{l=1}^M g_{l,k-1}^*(n)\Psi_{h_{m,k}}(n-l)   \label{eqn_24}
\end{align}
And also we have
\begin{align}
\Psi_{g_{m,k}}(n)&=\sum\limits_{l=1}^M a_{l,k-1}(n)\Psi_{g_{m,k}}(n-l)  \nonumber  \\
& \hspace{1.2cm}  +\sum\limits_{l=1}^M g_{l,k-1}(n)\Phi_{g_{m,k}}(n-l)                  \label{eqn_25}
\\
\Psi_{b_{m,k}}(n)&=\sum\limits_{l=1}^M a_{l,k-1}(n)\Psi_{b_{m,k}}(n-l)  \nonumber  \\
& \hspace{1.2cm}  +\sum\limits_{l=1}^M g_{l,k-1}(n)\Phi_{b_{m,k}}(n-l)                \label{eqn_26}
\\
\Psi_{h_{m,k}}(n)&=\sum\limits_{l=1}^M a_{l,k-1}(n)\Psi_{h_{m,k}}(n-l)  \nonumber  \\
& \hspace{1.2cm}  +\sum\limits_{l=1}^M g_{l,k-1}(n)\Phi_{h_{m,k}}(n-l)       \label{eqn_27}
\end{align}
In above relations, we need to update $4(M+N+1)$ sensitivities, but we can reduce the number of update relations to only eight update relations by using the approximation (\ref{eqn_17}). For instance, for term
\begin{align}{c}
\label{eqn_28}
\Phi_{a_k}(n)=[\Phi_{a_{1,k}}(n),\Phi_{a_{2,k}}(n),...,\Phi_{a_{M,k}}(n)]^T
\end{align}
We have
\begin{align}
\label{eqn_29}
\Phi_{a_{2,k}}(n)=y_k^* (n-2)+\sum\limits_{l=1}^M a_{l,k-1}^*(n)\Phi_{a_{2,k}}(n-l)\nonumber\\+\sum\limits_{l=1}^M g_{l,k-1}^*(n)\Psi_{a_{2,k}}(n-l)
\end{align}
That  can be replaced with
\setlength{\arraycolsep}{-0.07em}
\begin{eqnarray}
\label{eqn_30}
\Phi_{a_{2,k}}(n)&{}={}&\Phi_{a_{1,k}}(n-1)\nonumber\\&{}={}&y_k^* (n-1-1)+\sum\limits_{l=1}^M a_{l,k-1}^*(n-1)\Phi_{a_{1,k}}(n-1-l)\nonumber\\&&+\sum\limits_{l=1}^M g_{l,k-1}^*(n-1)\Psi_{a_{1,k}}(n-1-l)\nonumber\\&{}={}& y_k^* (n-2)+\sum\limits_{l=1}^M \underbrace{a_{l,k-1}^*(n-1)}_{\mbox{\footnotesize time delayed version}}\Phi_{a_{2,k}}(n-l)\nonumber\\&&{+}\:\sum\limits_{l=1}^M \underbrace{g_{l,k-1}^*(n-1)}_{\mbox{\footnotesize time delayed version}}\Psi_{a_{2,k}}(n-l)
\end{eqnarray}
This approximation can be applied to all other sensitivities. Finally, the update relations for weight vectors in the compact form is given by
\begin{equation}
\label{eqn_31}
\bw_k (n)=\bw_{k-1} (n)+\mu_k [e_k (n) \bPhi_{\bw_k} (n)+\bPsi_{\bw_k} (n) e_k^* (n)]
\end{equation}
This completes the derivation of the proposed IACA-IIR algorithm. 

\section{Simulations}
\begin{figure}[!t]
\centering
\includegraphics[width=9cm]{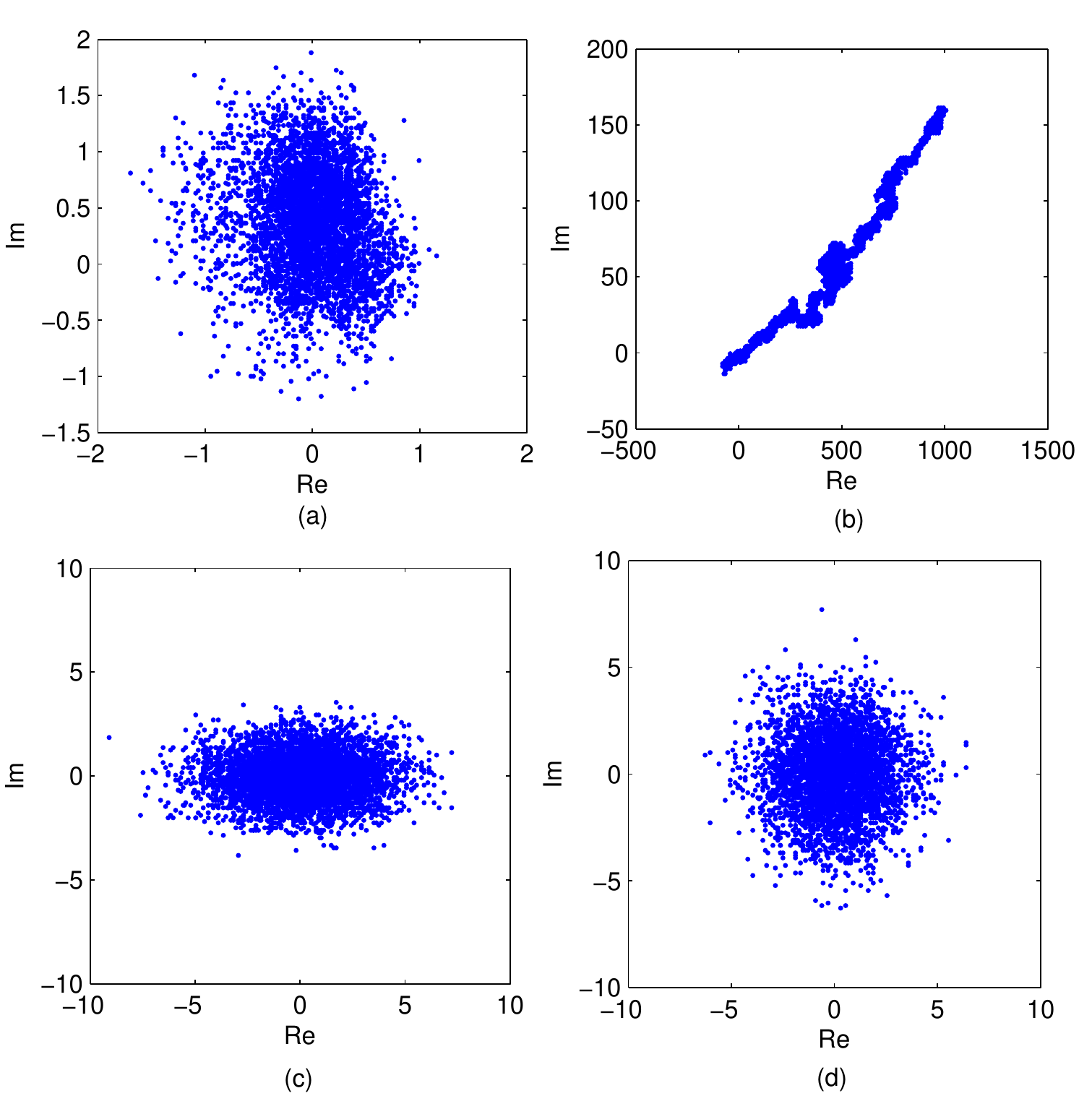}
\caption{Scatter plots for complex signals used in our simulations (a) circular complex–valued AR(4) process, (b) Proper MA process, (c) improper ARMA process, and (d) complex wind signal.}
\label{fig:2}
\end{figure}

In order to evaluate the proposed algorithm and to compare it with the non-cooperative case, we consider different complex-valued  signals in one-step Ahead prediction setting. One synthetic signal is a stable and circular
complex–valued AR(4) process used is given by
\begin{equation}
r(n) = 1.79r(n - 1) - 1.85r(n - 2) + 1.27r(n - 3)- 0.41r(n - 4) + z(n)
\end{equation}
where $z(n)$ denotes doubly white proper noise with unit variance. The other test signal is a linear MA which is a proper process which is described by \cite{19}:
\begin{eqnarray}
\label{eqn_32}
y(0)&{}={}&0 \nonumber\\
y(n)&{}={}&2z(n)+0.5z^* (n)+z(n-1)  \nonumber
\\&&+0.9z^* (n-1),n \geq 1
\end{eqnarray}
Other synthetic signal is a linear ARMA which is a improper process and is defined as a combination of  MA process (\ref{eqn_32}) and AR process, that is given by \cite{20}:
\begin{eqnarray}
\label{eqn_33}
r(0)&{}={}&0 \nonumber \\
r(n)&{}={}&1.8r(n-1)-1.85r(n-2) \nonumber \\
& & +1.3 r(n-3)-0.5 r(n-4)  \nonumber \\
&&+0.22 r(n-5)+2z(n)+0.5z^* (n) \nonumber\\&&+z(n-1)+0.9z^* (n-1),n\geq1
\end{eqnarray}
with
\begin{eqnarray}
\label{eqn_34}
\E{z(n-n_1) z^*(n-n_2)}=\delta(n_1-n_2) \nonumber \\
\E{z(n-n_1)z(n-n_2)}=\lambda \delta(n_1-n_2)
\end{eqnarray}
where $z(n)$ denotes doubly white proper noise, and $\lambda$ shows the noncircularity degree of signal. With $\lambda=0$, the proper MA model is achieved and with $\lambda=0.95$, the improper ARMA model is achieved. The third complex-valued signal is a real-world 2-D wind process\footnote{The wind data is available at \texttt{www.commsp.ee.ic.ac.uk/~mandic/wind-dataset.zip}} which is considered to have medium dynamic. 
Fig. \ref{fig:2} shows the scatter plots of the mentioned complex signals.
\begin{figure}[!t]
\centering
\includegraphics[width=8cm]{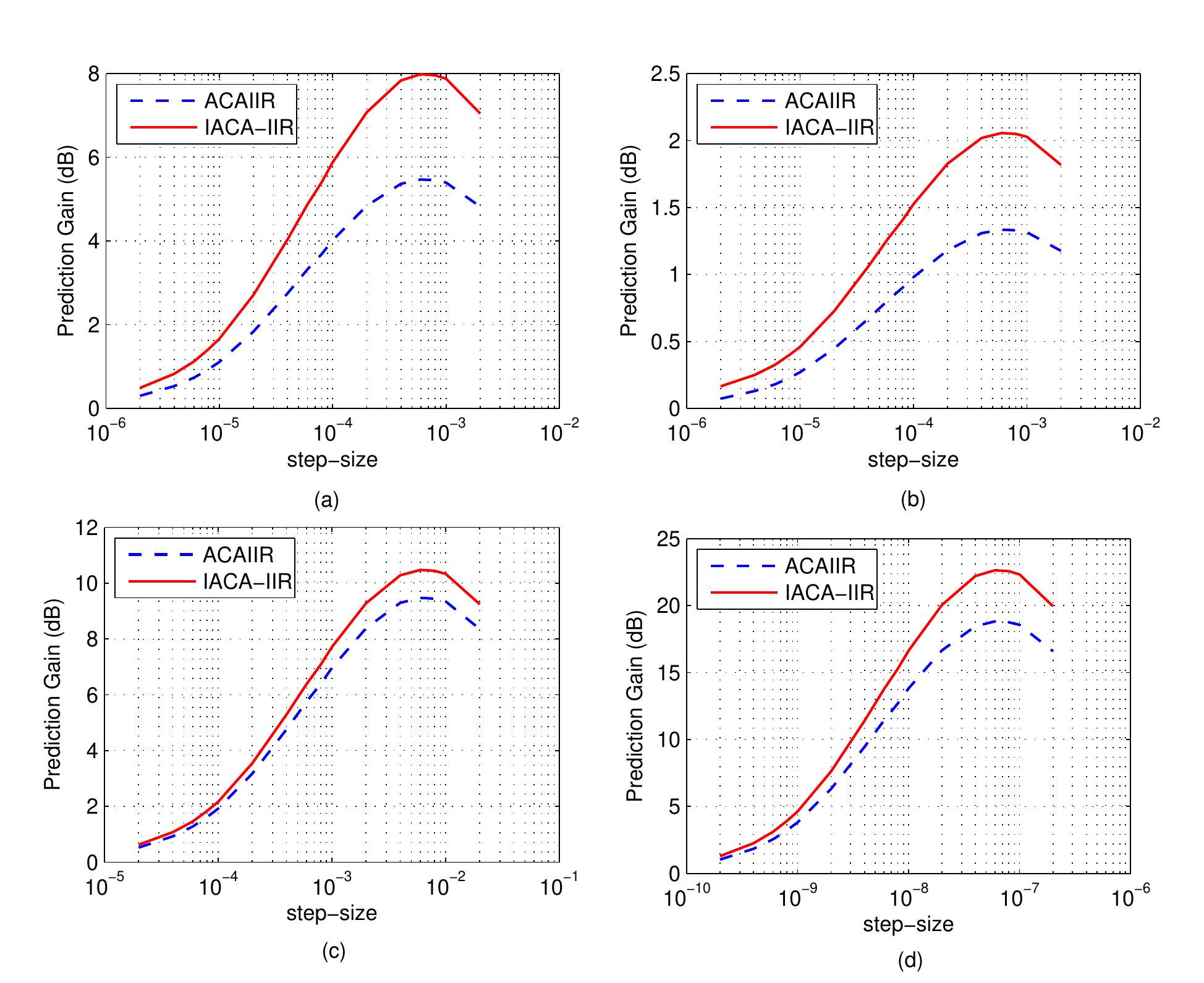}
\caption{Performance comparison between ACAIIR and IACA-IIR for prediction of (a) circular complex–valued AR(4) process, (b) Proper MA process, (c) improper ARMA process, and (d) complex wind signal.}
\label{fig:3}
\end{figure}

The network has $L=10$ nodes, and the filter order is set to $M=N=4$, that $M$ and $N$ denote respectively the order of feedback and feedforward of the filter. In order to evaluating the performance, the prediction gain $R_p=10\log(\sigma_x^2/\sigma_e^2)$  is used, where $\sigma_x^2$ and $\sigma_e^2$ are respectively the variance of the input and output error.

Fig. ~\ref{fig:3} illustrates the performance of proposed algorithm for prediction of complex-valued signals of interest and compares it with non-cooperative case over a range of step-size parameters. It can be observed that, in all situations, the proposed IACA-IIR algorithm outperforms the non-cooperative case. 
For a very small step size values, the adaptive algorithm can not track the signal (in prediction setting) and therefore the output variance is high (prediction gain is low). As step size increases, the ability of algorithm to follow the signal improves and the output variance decreases (prediction gain increases). Finally for large step sizes, the output error increases again and prediction gain decreases.

\begin{figure}[!h]
\centering
\includegraphics[width=8cm]{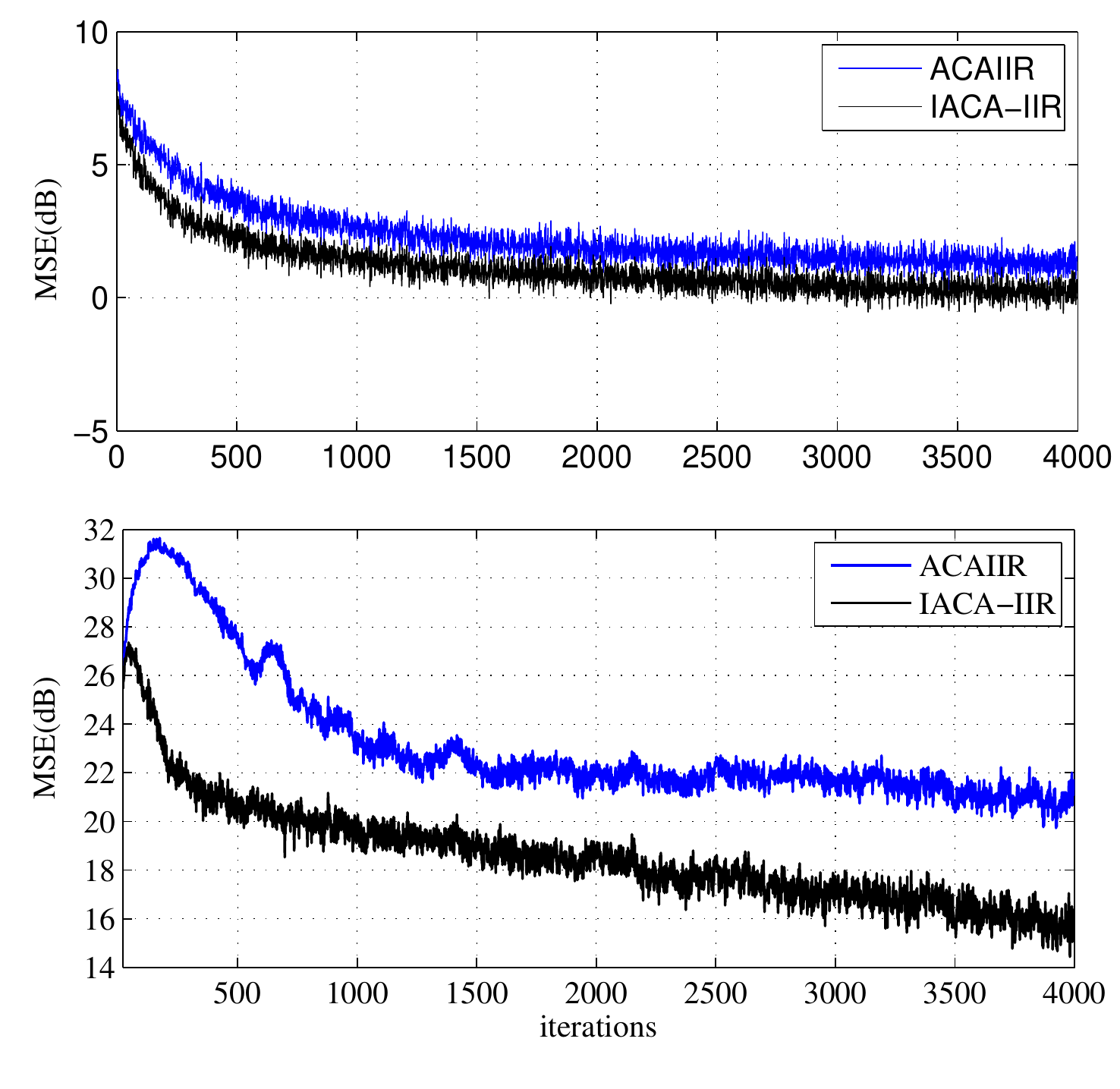}
\caption{Comparison of MSEs of ACAIIR  and IACA-IIR for prediction of (top) circular complex–valued AR(4) process, and (bottom) improper ARMA model.}
\label{fig:4}
\end{figure}

\begin{figure}[!t]
\centering
\includegraphics[width=8cm]{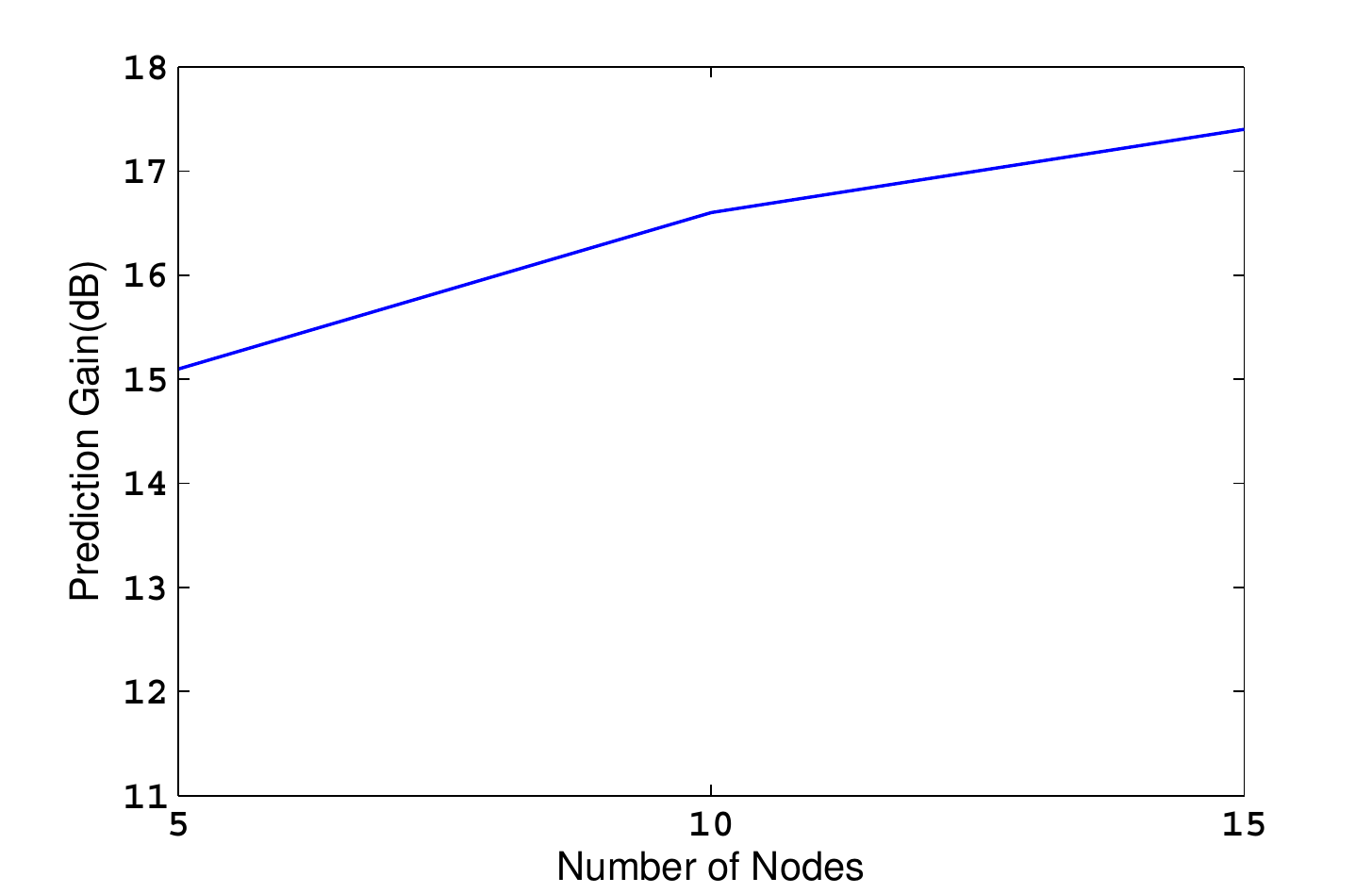}
\caption{The impact of network size on the performance of the proposed (in terms of the prediction gain) for wind signal.}
\label{fig:5}
\end{figure}

Fig. ~\ref{fig:4} illustrates the MSE curves of ACAIIR and IACA-IIR algorithms for prediction of circular complex–valued AR(4) signal, and improper ARMA model.  The step size parameters are set to $\mu=10^{-3}$ (for circular complex–valued AR(4) signal) and $\mu=10^{-7}$ (for improper ARMA model). It can be observed that, in this situation, the proposed IACA-IIR algorithm possesses faster convergence and smaller MSE than non-cooperative case.
The impact of network size on the performance of the proposed (in terms of the prediction gain) for wind signal (for $\mu=10^{-6}$) is shown in Fig. \ref{fig:5}. We can see that increasing the network size improves the performance of the proposed algorithm.

\section{Conclusion}
In this  paper we derive a distributed adaptive learning algorithm for training augmented complex adaptive IIR filters. The proposed algorithm (IACA-IIR) relies on the  augmented complex statistics and incremental cooperation mode. We used both synthetic and real-world circular and noncircular signals to evaluate the performance of the proposed algorithm, where the results reveal its superior performance over the non-cooperative ACAIIR algorithm. In our future work we plan to study the stability conditions for the proposed algorithm, especially in the different communications scenarios such as networks with noisy links.

%





\ifCLASSOPTIONcaptionsoff
  \newpage
\fi



\bibliographystyle{IEEEtran}
\end{document}